%% file: paper.tex
\documentclass[conference,anonymous]{IEEEtran}
\IEEEoverridecommandlockouts

\makeatletter
\renewcommand\footnoterule{%
  \kern-3\p@
  \hrule\@width.4\columnwidth
  \kern2.6\p@}
  \makeatother

\input{packages}

\input{macros}

\begin{document}

\title{Beware of the App! On the Vulnerability Surface of Smart
  Devices through their Companion Apps}

\author{
  \IEEEauthorblockN{Davino Mauro Junior}
\IEEEauthorblockA{Federal University of\\
  Pernambuco, Brazil\\
  dmtsj@cin.ufpe.br}
\and
\IEEEauthorblockN{Luis Melo}
\IEEEauthorblockA{Federal University of\\
  Pernambuco, Brazil\\
lhsm@cin.ufpe.br}
\and
\IEEEauthorblockN{Harvey Lu}
\IEEEauthorblockA{University of\\
  Michigan, USA\\
harveylu@umich.edu}
\and
\IEEEauthorblockN{Marcelo d'Amorim}
\IEEEauthorblockA{Federal University of\\
  Pernambuco, Brazil\\
damorim@cin.ufpe.br}
\and
\IEEEauthorblockN{Atul Prakash}
\IEEEauthorblockA{University of\\
  Michigan, USA\\
aprakashu@umich.edu}
}

\maketitle

\thispagestyle{plain}
\pagestyle{plain}


\begin{abstract}
Internet of Things (\iot{}) devices are becoming increasingly
important. These devices are often resource-limited, hindering
rigorous enforcement of security policies. Assessing the vulnerability of
\iot{} devices is an important problem, but analyzing their firmware is
difficult for a variety of reasons, including requiring the purchase of
devices. This paper finds that analyzing companion apps to these devices
for clues to security vulnerabilities can be an effective strategy.
Compared to device hardware and firmware, these
\app{}s are easy to download and analyze. A key finding of this
study is that the communication between an \iot{} device and its
\app{} is often not properly encrypted and authenticated and these
issues enable the construction of exploits to remotely control the devices. To confirm the vulnerabilities found, we created exploits against five popular IoT devices from Amazon by using a combination of static and
dynamic analyses. We also did a larger study, finding that analyzing
96 popular IoT devices only required analyzing 32 companion apps. Among
the conservative findings, 50\% of the apps corresponding to 38\% of the  devices did not use proper encryption techniques to secure device to companion app communication. Finally, we discuss defense strategies that developers can adapt
to address the lessons from our work.


\end{abstract}


\input{intro}

\input{problem}

\input{exploits}

\input{discussion}

\input{defense}

\input{related}

\input{conclusion}


\bibliographystyle{IEEEtran}
\bibliography{paper}

\end{document}

%% file: packages.tex
\usepackage[english]{babel}
\usepackage{textcomp}
\usepackage{float}
\usepackage{listings}
\usepackage{caption}
\usepackage{graphicx}
\usepackage{subcaption}
\usepackage{colortbl}
\usepackage{amsmath} 
\usepackage{mathpartir}
\usepackage{fancyvrb}\fvset{fontsize=\small}
\usepackage{xspace}
\usepackage[table]{xcolor}
\usepackage{balance}
\usepackage{wrapfig}
\usepackage{multirow}
\usepackage{pifont}
\usepackage{amssymb}
\usepackage{soul}
\usepackage{microtype}
\usepackage[T1]{fontenc}
\usepackage{dsfont}
\usepackage{relsize}
\usepackage{tikz}
\usepackage{longtable}
\usepackage{booktabs}
\usepackage{marvosym} 
\usepackage{framed}
\usepackage{mdwlist}
\usepackage{tabularx}
\usepackage{wrapfig}
\usepackage{array}
\usepackage{mathtools}
\usepackage{lscape}
\usepackage[linguistics]{forest}
\usepackage{afterpage}
\usepackage[shortlabels]{enumitem}

\usepackage{breakurl}
\usepackage[breaklinks]{hyperref}

\usepackage{pgf-pie}

%% file: macros.tex
\ifdefined\finalVersion
        \newcommand{\LineComment}[3]{}
        
\else
        \newcommand{\LineComment}[3]{\textbf{[[#1}{\color{#2}#3}\textbf{]]}}
        
\fi

\newcommand{\done}{$yes$}
\newcommand{\cross}{$no$}
\newcommand{\na}{$no\ encryption$}

\definecolor{darkblue}{rgb}{0.0, 0.0, 0.55}
\definecolor{cornellred}{rgb}{0.7, 0.11, 0.11}
\definecolor{gray}{rgb}{0.5,0.5,0.5}
\definecolor{purple}{rgb}{0.58,0,0.82}
\definecolor{bsgreen}{RGB}{104, 159, 56}

\newcommand{\Comment}[1]{}

\newcommand{\ie}{i.e.}

\newcommand{\eg}{e.g.}
\newcommand{\etal}{et al.}

\newcommand{\CodeIn}[1]{{\small\texttt{#1}}}
\newcommand{\Fix}[1]{\LineComment{}{red}{#1}}

\newcommand{\analyzedDevices}{five}

\newcommand{\totApps}{32} 

\newcommand{\totDevices}{96} 

\newcommand{\localComm}{18} 

\newcommand{\broadcast}{16}

\newcommand{\app}{app}

\newcommand{\iotapp}{IoT app}
\newcommand{\iot}{IoT}
\newcommand{\wifi}{Wi-Fi}
\newcommand{\smdevice}{smart device}

\newcommand{\qnone}{Q1}
\newcommand{\qntwo}{Q2}
\newcommand{\qnthree}{Q3}
\newcommand{\qnfour}{Q4}

\newcommand{\qone}{\emph{\qnone) Is the key hardcoded?}}
\newcommand{\qtwo}{\emph{\qntwo) Does the app use local communication?}}
\newcommand{\qthree}{\emph{\qnthree) Does the app send broadcast messages?}}
\newcommand{\qfour}{\emph{\qnfour) Does the app use any well-known protocol with vulnerabilities?}}

\newcommand{\kasavideo}{\url{https://figshare.com/s/d5bc439a7527df358f5f}}

%% file: intro.tex
\section{Introduction}\label{sec:intro}


The number of Internet of Things (\iot{}) devices~\footnote{We may
  refer to \iot{} devices as smart devices (or just devices) and the
  \app{}s that control these devices as companion \app{}s (or just
  \app{}).}  worldwide is predicted to reach 20 billion by
2020~\cite{gartner-20billion-devices-2020}. Designing secure solutions
in this domain is challenging as devices are typically limited in
resources. Consequently, security is a permanent concern. As a
concrete example, in October 2016, the Mirai malware compromised
millions of IoT devices around the world and used them to launch the
largest DDoS attack ever recorded~\cite{icit-rise-2016}. In the
smart-home scenario, security vulnerabilities in IoT devices could
compromise safety at home~\cite{denning-kohno-levy-2013}.

\iot\ devices are compatible with multiple cloud-based software stacks
(\eg{}, SmartApps in the SmartThings cloud~\cite{smartthings}, Alexa
Skills~\cite{echo-alexa}, etc.). Prior work has found security
vulnerabilities introduced by some of these
stacks~\cite{fernandes-jung-prakash-2016,DBLP:journals/corr/abs-1712-03327}. Unfortunately,
a gap remains in that the device may have vulnerabilities
out-of-the-box that are independent of security of high-level software
stacks. Unfortunately,
techniques for security analysis of device software itself remains an
art and poorly understood. For example, a white paper by Veracode
describes vulnerabilities in 6 IoT devices, including SmartThings Hub,
Wink Hub, Wink Relay, and MyQ Garage, but not how to analyze such
vulnerabilities in a systematic way.  One solution would be for a
security analyst to inspect the firmware binary on the device, but
that is often hard to access~\cite{DBLP:conf/ndss/ChenWBE16} and
challenging to analyze~\cite{cmu-bap}.

This paper proposes an indirect and simpler way of assessing the
security of \iot\ devices by analyzing their companion apps and the
interaction with the device's firmware. Our intuition is that if this
interaction between the companion app and device firmware is not
implemented with good security principles, the device's firmware is
potentially insecure and vulnerable to attacks. In our experience,
most IoT devices on the market today are released with companion apps
for both Android and iOS so that users could control these devices
directly from their smartphone, thus permitting such analysis.

Our hypothesis is that the analysis of these apps can throw
substantial light on potential vulnerabilities in devices and even
help security analysts develop proof-of-concept exploits to induce the
device manufacturers to verify the vulnerabilities and fix them.  To
validate the hypothesis, we analyze multiple smartphone \app{}s to
discover potential vulnerabilities and, from that, create
proof-of-concept attacks that could allow either a local or remote
attacker to completely compromise the device, including issuing
arbitrary commands to them or update their firmware without ever
touching the device physically. These \app{}s play an important role in
the secure operation of smart devices. In particular, they are
responsible for the initial device configuration, \ie{}, these \app{}s
set up the communication channel with the device with supposedly
proper encryption and authentication. Because of this important role,
\iotapp{}s need to encode sensitive communication data, such as the
type of encryption being used, and even encryption keys. This
information can be valuable for an attacker who wants to gain control
of the device. Moreover, in contrast to physical devices, these
\app{}s are accessible to the general public through market places
(\eg{}, Google Play). It is, therefore, a sweet spot for hackers to
focus their efforts on.


In the security analysis of the smartphone \app{}s, we particularly
focus on the security of communication between an \iot{} device and
its \app{}. Any flaws could enable a range of attacks that can result
in complete control of the IoT device\Comment{ (see
  Section~\ref{sec:threat-model})}. We analyzed smartphone \app{}s
for 96 of top-selling WiFi and Bluetooth-enabled devices on Amazon.
There were 32 unique apps for these devices (e.g., devices were from the
same vendor and sometimes different vendors shared apps) and then
developed proof-of-concept attacks on 5 of the devices.  For example, we find that an Amazon top-seller smart plug
from TP-Link~\cite{tp-link} shares the same hard-coded encryption key
for all the devices of a given product line and that the initial
configuration of the device is established through the \app{} without
proper authentication. Using this information, we were able to create
a spoofing attack to gain control of this device\footnote{A video illustrating a counterfeit app in action can be
  found at \kasavideo{}}. Note that this issue can be replicated in
all other TP-Link devices that use the same app.\Comment{ In total, we found that
  \Fix{N} devices use the same app.}

This paper makes the following contributions:

\textbf{Empirical Study.}~We show the value of analyzing companion apps
as a useful vulnerability analysis technique for devices by conducting two studies to assess the security
of app-device communication by analyzing companion apps corresponding to the device. These studies consider key
security aspects in this context, namely, encryption, authentication,
and network protocols. The first study involved a detailed analysis of four different apps (for five devices, with two devices sharing an app) and
then creating concrete exploits. A second study analyzed 32 apps for 96 devices
to find extent of similar features that could potentially enable an exploit.

\textbf{Efficiency.} We found that we only had to analyze 4 apps for the 5
devices actually purchased and 32 apps for 96 devices overall. There was extensive sharing of apps among devices. Thus, it can be significantly more efficient
to analyze companion apps as compared to device firmware.


\textbf{Findings and Lessons.}~We found lack of encryption in 31\% of
the apps analyzed and use of hardcoded keys in 19\% of all the apps---thus,
at least 50\% of the apps were potentially seriously vulnerable to exploits
(these apps corresponded to 37 out of 96 devices considered).
Many of these apps controlled their devices via local communication or via
broadcast messages, including UDP messages. Based on our in-depth analysis
of 4 of the apps, we found that leveraging these weaknesses to create
actual exploits is not challenging. A remote attacker simply has to find
a way of getting the exploit either on the user's smartphone in the form
of an unprivileged app or a script on the local network. We then
discuss potential defense strategies.

\textbf{Spoofing Attacks.}~We provide detailed evidence of the
importance of our findings by building proof-of-concept attacks
on randomly selected, from
a list of popular IoT devices on Amazon, \analyzedDevices{}
IoT devices. We purchased these devices prior to the analysis.
We were successful in creating attack \app{}s (or scripts)
that could execute arbitrary commands on the device.
The attacks were successful even if the device
had previously been paired with a legitimate app.

%% file: problem.tex
\section{Context, Goal, and Questions}\label{sec:context-problem}

\begin{figure}[t]
  \centering
  \includegraphics[trim=0 120 0 0,clip,width=0.5\textwidth]{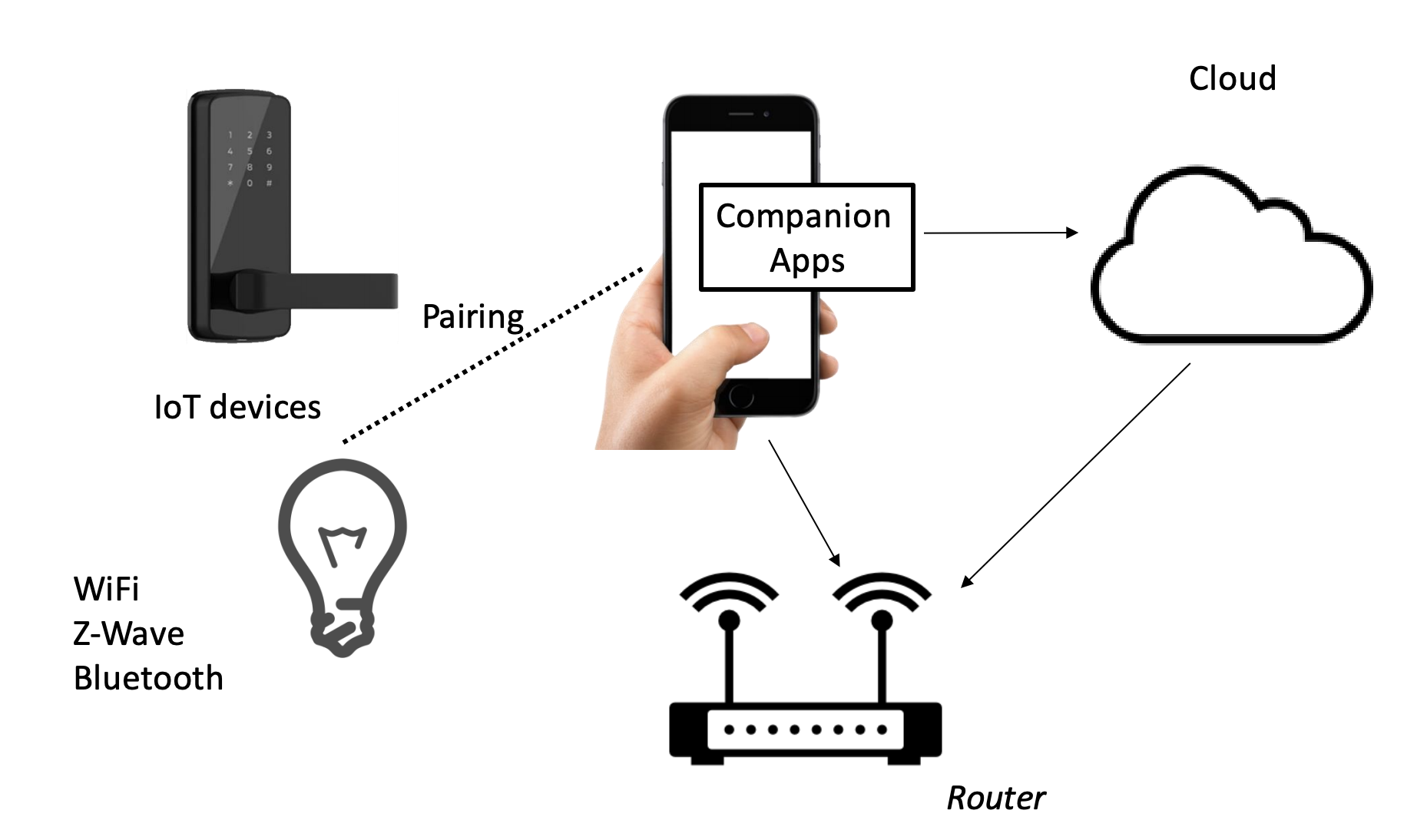}
  \caption{\label{fig:iot-setup}Example of \iot\ setup.}
\end{figure}

\subsection{Context}
Most manufacturers of IoT devices provide a smartphone
application and cloud services to monitor and control their
devices. Communication between the device and its companion \app{} is
often established over the local network using some
wireless protocol such as Zigbee and
Z-Wave~\cite{the-ambient-zigbee-zwave}. When the device and the smartphone are in distinct networks
or the \iot{} setup does not support local network
connectivity\footnote{Tuya Smart~\cite{tuyasmart} is an example
  framework where the companion \app{} can only communicate with the device
  through the cloud.}, the cloud server
acts as a proxy for communication. An \iotapp{} sends the message
intended for a device to the cloud server and the cloud server relays
the message to the device\Comment{ through the hub}. Similarly, a message from a device can be
relayed to the \iotapp{} through the cloud. Figure~\ref{fig:iot-setup}
illustrates an example setup with a router/hub and with the
cloud. Setups without one or both of these elements are possible. For
example, it is possible that only
local communication is allowed (\ie, no cloud) and that the \app\ and
the device communicate directly (\ie{}, through the router). Hubs are
especially useful to enable communication with resource-constrained
devices that do not need to fully implement a network stack.

In this context, pairing is the process of establishing a communication channel between
an app and a device. A prevalent hypothesis made by users is
that, once established, this channel is secure. The goal of this paper
is to assess this hypothesis, \ie{}, this study evaluates whether the
app-device channel is properly secured. To
achieve this goal we analyze companion apps for
vulnerabilities and then confirm them by creating exploits. We make the following assumptions
on these exploits:
\begin{itemize}
  \item that an adversary has access to the local network;\Comment{ Remote
    communication through the cloud is out of the scope of this
    study.}
  \item that an adversary uses a rogue app or script to control the
    device without user knowledge.
\end{itemize}


To illustrate a potential use case for such exploits, consider, for
instance, the scenario of a burglar trying to break into your smart
home. The first step is to gain access to your WiFi network using some
public technique~\cite{vpn-guru-wifi-hacking}. The burglar does not
need to succeed in all attempts. The next step is to detect periods
of time when the house would be empty. For that, the burglar could monitor the
network and find use-patterns to identify periods of
occupation. Finally, the burglar would use a rogue app or script to
control a door handle and invade the house. This scenario illustrates
the importance of assessing whether the \iot\ device is susceptible to
adversaries leveraging information obtained by analyzing the companion
app.

\subsection{Goal and Questions}\label{sec:questions}

The goal of this study is to assess whether the app-device communication is secure.
We assume this communication occurs through the local network;
remote communication through the cloud is out of the scope for this
work. We also assume that
an adversary targeting the \iot\ device would use either a rogue app
or script to control the device without user knowledge.
We propose a non-comprehensive list of questions to determine the
attack surface on the apps.\Comment{addressing the
goal of assessing how secure is the app-device communication channel.}
These questions are related to encryption, authentication, and
communication, which is where vulnerabilities typically manifest.


\qone{}~A malicious developer
could counterfeit messages if she has access to secret keys. Hardcoded
keys are problematic as they are encoded literally in code\Comment{ for
securing a communication channel}. Intuitively, hardcoded keys could be
mined by reverse-engineering the \app{}, even when the code is
obfuscated. For example, if a crypto API is used, one could find the
key by monitoring the actual parameters of crypto library functions,
whose names cannot be obfuscated and whose intent can be found from
public documentation.  In other cases, the developer may have chosen
to implement a custom crypto function, but one could discover these
functions by the ratio of mathematical instructions in the function
body, as Caballero and colleagues
did~\cite{caballero-poosankam-kreibich-song-2009}, and then use a
similar method to mine the key.


\qtwo{}~When the \iotapp{} and the corresponding device
are in the same network, local communication may be
used. Unfortunately, local communication protocols do not enforce the same security guarantees compared to
remote communication. For example, it is uncommon to check identities
with HTTPS/SSL certificates in local communication. In contrast, in
the scenario where local communication is forbidden (\ie{}, cloud
relays messages to \app{}s and devices), an attacker would have to
make a bigger effort to forge HTTPS/SSL certificates of the parties
involved\Comment{--the device and the cloud server}.


\qthree{}~Broadcast messages are frequently used in IoT setups to
discover devices and to enable direct app-device communication when
there is no hub/gateway in the setup. Their use, unfortunately, can
put a smart home at risk.  Adversaries can, for instance,
sniff\footnote{This ability to sniff WiFi messages depends on the
  distance to the router.} the response of devices to broadcast
messages, which often include sensitive data such as the internal
state of the device.

\begin{table}[t]
  \centering
  \caption{\label{table:cve}CVE vulnerabilities in major \iot{} protocols.}
  \begin{tabular}{lrc}
    \toprule
    Protocol & \# Vulnerabilities & Example \\
    \midrule
    MQTT & 13 & \href{https://cve.mitre.org/cgi-bin/cvename.cgi?name=CVE-2017-9868}{CVE-2017-9868}\\
    SIP & 59 & \href{https://cve.mitre.org/cgi-bin/cvename.cgi?name=CVE-2018-0332}{CVE-2018-0332}\\
    UPnP & 346 & \href{https://cve.mitre.org/cgi-bin/cvename.cgi?name=CVE-2016-6255}{CVE-2016-6255}\\
    SSDP & 17 &  \href{https://cve.mitre.org/cgi-bin/cvename.cgi?name=CVE-2017-5042}{CVE-2017-5042}\\
    \bottomrule
  \end{tabular}
\end{table}

\qfour{}~Different protocols tailored to \iot{} deployments exist and
some of these protocols are known to be vulnerable to
attacks. According to Al-Fuqaha and colleagues, a total of seven
application protocols are more frequently used in \iot{}
deployments~\cite{iot-survey-2015}.  Using the Common Vulnerabilities
and Exposures (CVE) database~\cite{cve}, we find vulnerability
reports in four of these \iot{} protocols, namely, MQTT~\cite{mqtt},
SIP~\cite{sip}, UPnP~\cite{upnp} and SSDP~\cite{ssdp-draft}.\@
\autoref{table:cve} shows the number of reported issues and the ID of
an example issue. For instance, the UPnP vulnerability
\href{https://cve.mitre.org/cgi-bin/cvename.cgi?name=CVE-2016-6255}{CVE-2016-6255}
allows remote attackers to write arbitrary files to the device
file system~\cite{upnp-flaws}. Note that UPnP is the protocol with the
largest number of reported issues.

%% file: exploits.tex
\section{Finding and Confirming Vulnerabilities}\label{sec:threat-model}\label{sec:finding-confirming}

This section presents details of how we carried out a vulnerability assessment
by analyzing the companion apps and then confirming the assessment by crafting proof-of-concept
exploits for a selection of IoT devices. It describes the criterion for selecting apps to
analyze~(\ref{sec:app-selection-criteria}), discusses the analysis used to answer the questions
posed on Sections~\ref{sec:questions} and~\ref{sec:exploits-analysis},
and describes each exploit in detail~(\ref{sec:exploits}).

\subsection{App Selection Criterion}\label{sec:app-selection-criteria}
To select \app{}s for this part of our study, we examined the 96
top-selling WiFi and Bluetooth devices on the Amazon website by
popularity. We then restricted the resulting set to devices from the
categories smart plugs, bulbs, or IR controllers that use
\wifi{}---largely for affordability and form-factor reasons and
because \wifi{} is popular and provides a potential attack surface if
an attacker can execute code anywhere on the same network (e.g., via
an app, malicious email, malicious device, or downloaded executable
code on a computer on the same network). A total of 54 devices
satisfied this criterion. From these 54 devices, we randomly selected
(and purchased) 5 devices to run our analysis. Somewhat to our
surprise, we found that two of the devices we selected use the same
app (as they belong to the same manufacturer).  Consequently, this
section focuses on these four companion apps. For each of the apps, we
did a detailed vulnerability analysis and then also developed
exploits.

\subsection{Vulnerability Analysis}\label{sec:exploits-analysis}
This section details the tools we used to determine app-device protocol
features that could permit remote or local attacks on the device.
We analyzed each companion app with respect to the questions from
Section~\ref{sec:questions}. We then identified a potential attack path
in each app and confirmed the path by creating a proof-of-concept exploit.
We describe the methodology used to answer the questions and their use
to find an exploit.

\subsubsection{Basic toolset functionality}~We implemented a toolset
to help us do semi-automated analysis to answer the
questions\Comment{ from Section~\ref{sec:questions}}. We found relying
on just automated analysis to be error-prone.  For example, one of our
tools looked for use of constant keys in calls to encryption
functions. But, we found that some of the calls were not used on the
communication paths between the device and the app.  Thus, for all of
our analysis, we used the tools as an aid to manual analysis.  We also
used the JADX decompiler library~\cite{jadx} as well as some static analysis tools~\cite{amandroid}.

\vspace{0.5ex}\noindent\textbf{Encryption Discovery.}~The encryption
discovery component looks for functions in the \app{} that likely
encrypt and decrypt the data exchanged with the \smdevice{}. Those
functions are the first line of attack for
adversaries~\cite{crypto-api-misuse}. With those functions, one could,
for example, eavesdrop on official communication and infer the layout of
messages and gain access to sensitive data.  The toolset uses two
complementary heuristics to discover these encryption functions. The
first heuristic applies to the case where developers use existing Java
encryption APIs. The second heuristic covers the case where developers
implement custom crypto functions instead of building on existing
ones. The toolset detects these functions by computing, for every
function declared in the \app{}, the ratio between the number of
arithmetic and bitwise operations over the total number of
instructions. This heuristic has been previously used in prior
work~\cite{caballero-poosankam-kreibich-song-2009,reformat-2009,iotfuzzer-2018}. \\
\noindent\textbf{Network Protocol Discovery.}~This component extracts
information about the communication protocol used between the
\smdevice{} and the companion \app{}. More precisely, it looks for
calls to functions (in the \app{}) from classes related to known
communication protocols. For example, for UDP, it looks for calls to
functions from \CodeIn{java.net.DatagramSocket} and, for TCP, it looks
for calls to functions from \CodeIn{java.net.Socket}. This component
reports a mapping from classes in the \app{} to communication
protocols (\eg{}, TCP, UDP, HTTP, UPnP). Note
that\Comment{, although our current solution is specific to Java and
  these protocols,} there is no fundamental limitation that prevents
our infrastructure supporting other languages and protocols.

\subsubsection{Answering the questions}



We describe below the methodology we used to answer each of the questions from
Section~\ref{sec:questions} and then elaborate results.

\qone{}~The search for hardcoded keys initiates from the output of the encryption discovery
component, which reports function likely related to encryption. When using
standard encryption libraries, we are able to automate the
search for secret keys by looking for \CodeIn{javax.crypto.SecretKey},
which is the class denoting a
key in the Java standard API\@. For custom encryption, however, we
manually inspect each method returned by the encryption discovery,
checking if the key is present inside the method body or in usages of the method.

\qtwo{}~The protocol discovery component acts as guidance for
the manual analysis. Based on the function calls and protocol report,
we manually analyze the classes responsible for network calls and
identify whether the app uses local communication.


\qthree{}~Identifying whether broadcast messages are sent
from the \app{} to the \smdevice{} is done by inspecting the classes
responsible for making network calls and looking for well-known
broadcast addresses, \eg{}, \CodeIn{255.255.255.255}.

\qfour{}~With the protocol discovery component, we can
identify transport, network, and application layer protocols used on
the \app{}. After identifying these protocols, we look for documented
vulnerabilities for each protocol in the Common Vulnerabilities and
Exposures (CVE) database~\cite{cve}.

\begin{table*}[!t]
  \centering
  \caption{Potential Threats to Selected Apps.}
  \input{tables/appFeatureExploits}\label{table:appFeatureExploits}
\end{table*}

\vspace{0.5ex}\noindent\textbf{Results.}~Table~\ref{table:appFeatureExploits} shows the answer to these
questions for the four selected apps. For each question, we used the
labels \done{} or \cross{} to indicate a positive answer or a
negative answer, and \na{} for the first question when the app uses no encryption. The
label \done{} indicate good practice whereas the labels \cross{} and \na{}
indicates a potential vulnerability or an interesting attack surface for
a potential exploit.


All four apps are found to use local communication with
the device and three of the apps also using broadcast communication,
providing us with a potential attack surface to exploit the devices.
Three out of four apps do not use any encryption to secure their
communication with the device, providing us with a compelling attack vector.
Only one of the selected apps (WeMo) uses an insecure version of a protocol.
But WeMo also does not use any encryption, thus providing a simpler
attack vector to exploit the device.

\subsubsection{Finding Vulnerable Paths}\label{sec:vulnerable-paths}

After answering the posed questions, we proceed to locate {\em vulnerable
  paths} in the given app. We define a vulnerable path to be a sequence of
function calls that connect a source (\eg{}, a
function that is called from external input such as the user interface (UI)) to a sink
(\eg{}, a network method call) that an exploit may wish to compromise.
Analyzing the classes and functions in
this path involved in the preparation of a message to a sink helps the security
analyst generate an abstraction of the application behavior, an important step to
creating an exploit.


\begin{figure}[h!]
  \centering
  \includegraphics[width=0.4\textwidth]{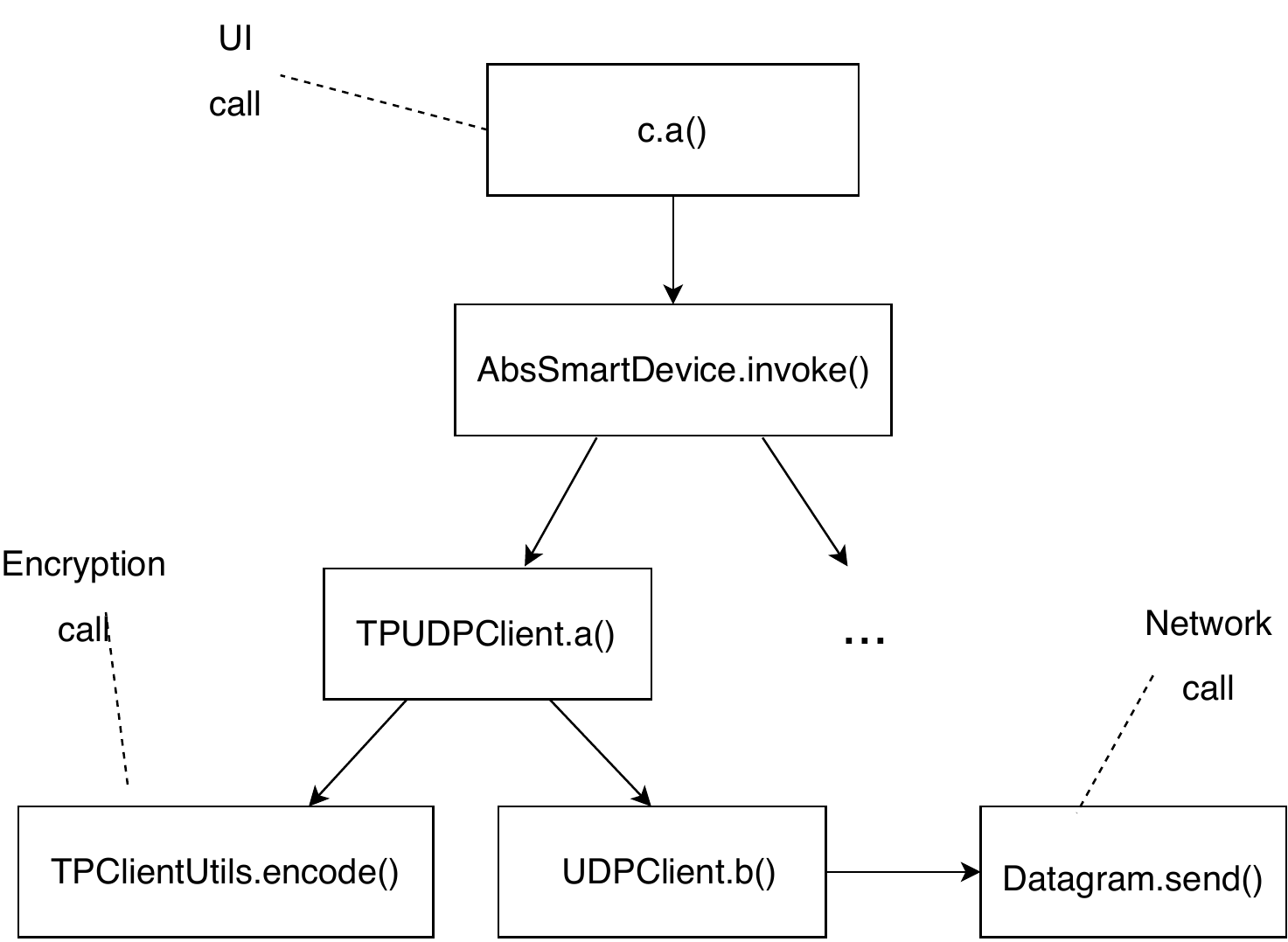}
  \caption{\label{fig:call-kasa}Path (simplified) from UI function to a network call.}
\end{figure}

Figure~\ref{fig:call-kasa} illustrates a vulnerable path for
the Kasa app. To find this vulnerable path, we start by analyzing the output of our
toolset, \ie{}, classes and functions related to encryption,
authentication and network protocols. These elements are potential sources of
vulnerabilities. Considering the Kasa app, for example, we
start by inspecting the classes containing
usages of the UDP protocol (related to \qnthree{}). We discover that the \CodeIn{UDPClient}
class declares the network-related method \CodeIn{b}, which calls
\CodeIn{datagramPacket.send()}, a method from
the standard Java API to send UDP packets. As the method
\CodeIn{b} includes a network call, it could be flagged as a sink. Our analysis shows that this class contains usages of
broadcast addresses, representing a potential attack surface. Then, we analyze backward the call chain leading to this
function looking for a UI method. While doing it, we find another method
present in the output of our
toolset (related to \qnone{}), \CodeIn{TPClientUtils.encode}, contains hardcoded keys
that could also be exploited. We also identify the function \CodeIn{TPUDPCLient.a}, responsible
for building the UDP packet. This
function, while not showing a vulnerability by itself, is responsible for
building the UDP packet to be sent and reveals
the structure of the message. Finally, we discover the calls to the
UI obfuscated method \CodeIn{c.a}, which is the
starting point of this path. That is possible because of the
programming conventions of Android. More specifically, class \CodeIn{c}
declares several (button-related) event callback
methods.

\subsection{Exploits}\label{sec:exploits}

Based on the vulnerable paths found, we created exploits for each
selected app. In the following, we describe in detail the steps
we took to create each exploit. Although this work does not assume
that an adversary would have physical access to the device, we used
physical devices in this experiment to demonstrate our findings. The
assumption we made on the adversary is that she understands Java and
she is able to reverse engineer the Android application files
(.apks). Also, when cryptographic functions and keys are encountered,
we do not assume the adversary has enough computational power to break
the key by brute force. We believe that these assumptions are similar
to what a real-world attacker would deal with and similar to the
assumptions made in other studies~\cite{auto-hacking, bank-hacking}.

\input{exploit_kasa}

\input{exploit_lifx}

\input{exploit_wemo}

\input{exploit_econtrol}

\subsection{Vulnerability Disclosure} As of the publication deadline of
this paper, we have notified all manufacturers of the
vulnerabilities. During disclosure, we also included scripts showing how to control
the device without the official \iotapp{} and possible approaches to
mitigate the issues we found. None of them have sent any
response to our disclosures and to the best of our knowledge, have not
released patches relative to these
vulnerabilities.

\subsection{Threats to Validity and Limitations}~The main threats to the validity of this study are the
following. \textit{External Validity:}~We ran this study against five
devices, associated with four different \app{}s. Although we analyzed
more apps (see Section~\ref{sec:discussion}), we only purchased these
devices. As usual, the results may not generalize to other devices. To
reduce bias, we selected smart home devices according to a
well-defined criterion based on popularity in the Amazon website. It is worth noting that the characteristics of the \app{}s that
control the devices we purchased are similar to the ones we
analyzed but not purchased.\\\textit{Internal Validity:}~Our
results could be influenced by unintentional mistakes during human
inspection. For example, we could have missed an important vulnerable
path in an app.

As for limitations, our infrastructure and methodology do not
account for the possibility of \app{}s making Java Native Interface
(JNI) calls. In this study, none of the apps calls native functions
through JNI\@. In a complementary study (see
Section~\ref{sec:discussion}), however, we found that some apps make
JNI calls. They use JNI, for example, to implement encryption. As
expected, reverse-engineering native code is more challenging.



%% file: tables/appFeatureExploits.tex
\definecolor{lightgray}{gray}{0.94}

\setlength{\tabcolsep}{1pt}
\begin{center}
  \begin{footnotesize}
    \begin{tabular}{lccccccc}
        \toprule
        \multicolumn{1}{c}{\textbf{App}} 
      & \multicolumn{1}{p{2.2cm}}{\centering\textbf{Avoid Hardcoded Keys?}} 
      & \multicolumn{1}{p{2.2cm}}{\centering\textbf{Avoid Local Communication?}}
      & \multicolumn{1}{p{2cm}}{\centering\textbf{Avoid Broadcast Messages?}}
      & \multicolumn{1}{p{2cm}}{\centering \textbf{Safe}\\ \textbf{Protocol?}}\\
      \midrule
      Kasa for Mobile	& \cross{}	& \cross{}	&\cross{}	& \done{}\\
      LIFX              & \na{}	        & \cross{}	& \cross{}      & \done{}\\
      WeMo	        & \na{}	        & \cross{}	& \done{}	& \cross{}\\
      e-Control         & \na{}	        & \cross{}	& \cross{}	& \done{}\\
      \bottomrule
      \end{tabular}
\end{footnotesize}
\end{center}

%% file: exploit_kasa.tex
\subsubsection{Finding an exploit for Kasa for Mobile}\label{kasa-exploit}

TP-Link Kasa is the official app for controlling
TP-Link-manufactured devices from the smart home product line
Kasa~\cite{tp-link-kasa}. The exploit we created consists of a rogue
app that mimics the official TP-Link app and takes control of a
TP-Link smart plug. It is worth noting that in principle, this app
could run as code anywhere on the same network, \eg{}, running as a
script instead of app. At the time of this writing, TP-Link's
smart plug was a top-seller with over
12.000 customer reviews on the Amazon
website~\cite{tp-link-plug-amazon} showing an average rating of 4.4 out
of 5 stars. We give an analysis of its companion app below.

\qone{}~The Kasa app uses a custom encryption function, Caesar
cipher~\cite{caesar-cipher}, that is known to be easy to
break. Listing~\ref{lst:kasa-encryption-function} shows this
function as it appears in the app. Line 2 shows the hardcoded seed to encrypt the data. Identifying the encryption function and its hardcoded seed
gave us hope of replicating the function in a rogue app on the same
network to control the device arbitrarily.

\begin{lstlisting}[float,language=Java,
                 label={lst:kasa-encryption-function},
                 caption={TP-Link Kasa encryption
                 function.}
                ]
public static byte[] encode(byte[] data) {
  byte seed = (byte) -85;
  for (int i = 0; i < data.length; i++) {
    data[i] = (byte) (data[i] ^ seed);
    seed = data[i];
  }
  return data;
}
\end{lstlisting}

\qtwo{}~By using the network discovery component and manually inspecting the code, we identified
classes containing calls to UDP-related methods. After inspecting these classes, we
confirmed that these methods are involved in the discovery and
control of the TP-Link devices on the local network. For instance,
Listing~\ref{lst:kasa-discovery-function} exhibits the
function that discovers TP-Link devices in the local network.

\begin{lstlisting}[language=Java,
                 escapechar=@,
                 label={lst:kasa-discovery-function},
                 caption={TP-Link Kasa function (simplified) used to
                 discover devices on the local network.}
                ]
public static final String UDP_ADDRESS="255.255.255.255";@\label{line-udp-address-def}@
public void discoverLocal() {
  String requestId = DiscoveryUtils.a();
  tpDiscovery.@\textbf{broadcastDiscovery}@(...,UDP_ADDRESS,...); @\label{line-udp-address-use}@
  ...
}
\end{lstlisting}


\qthree{}~During our analysis, we found the Kasa app uses broadcast
                 messages to discover and control the TP-Link devices.
                 Line~\ref{line-udp-address-def} from
Listing~\ref{lst:kasa-discovery-function} declares a constant variable
holding a well-known broadcast IPv4 address. This variable is then
used in Line~\ref{line-udp-address-use} to discover TP-Link devices on
the network.

\qfour{}~Our analysis did not find usage of protocols
with documented vulnerabilities.


To confirm the vulnerabilities, we designed a proof-of-concept exploit. The
exploit consists of a rogue app on the same network.
We use both static
analysis (\ie{}, inspection of the decompiled code) and dynamic
analysis (\ie{}, inspection of the network traffic). The list below
shows the keys steps to create the exploit.
\begin{enumerate}
\item\label{step:a}\label{step:b} find a vulnerable path and encryption function;
\item\label{step:c} discover the structure of exchanged messages;
\item\label{step:d} discover what protocol is used to exchange messages;
\item\label{step:e} implement pairing.
\end{enumerate}

From one vulnerable path for the Kasa app (see
Section~\ref{sec:vulnerable-paths}), we obtain access to the app's
encryption function (step~\ref{step:a}). With that function at hand,
it is possible to monitor the network traffic and read the contents of
messages as to understand their structure (step~\ref{step:c}) and the
IP addresses used (step~\ref{step:d}), which are critical for
replication. Recall that the Caesar cipher~\cite{caesar-cipher} is a
symmetric (/invertible) encryption function. In that case, we found
that only broadcasting was used through a single address.  Finally, it
is necessary to replicate pairing (step~\ref{step:e}). To our surprise, we found by inspection that a pairing process was
 not needed to control the device. Pairing was used to
 maintain the profile of users on TP-Link devices, but not for its control.



\vspace{1ex}\noindent\textbf{Monitoring the Network.~} We
used the popular traffic analyzer Wireshark~\cite{wireshark} to
monitor the packets exchanged between the Kasa app and the
device. As the traffic was encrypted we needed to implement a script
to decrypt the monitored messages; the script uses the symmetric
cipher function from Listing~\ref{lst:kasa-encryption-function}. This monitoring tool was used in two important stages:~(i) during the
app-device pairing process and (ii) while the app interacted with the
device, \eg{}, turning the plug ``on'' and ``off''. During the pairing
process, we found that broadcast messages were exchanged while the app
was connected to the hotspot created by the device. We also monitored the network
when interacting with the device through the app's UI\@. Specifically,
we repeated the ``Turn Off'' and ``Turn On'' operation multiple times,
observing that the contents of the network packets did not change,
validating the use of a hardcoded key with a poor encryption method.
We also observed the use of broadcast messages during device usage after pairing. We found that the app uses the following message to discover and obtain
the current status of the
device--\CodeIn{\{"system":\{"get\_sysinfo": \}\{\}\}}. We also found
that \CodeIn{"\{"system":\{"set\_relay\_state":\{"state":0\}\}\}}
was the message for turning the device off.

Based on the analysis, we created a rogue app to control a TP-Link smart plug
device. Static analysis played an important role to find vulnerable
paths in the app whereas dynamic analysis helped in understanding
the communication protocol and the messages exchanged. Recall that, during our
analysis, we noticed that the pairing process was not needed to control
the device. This is a severe flaw as the user would not
even be aware of an attack---the official app would still work as
intended even with a rogue app controlling the device
simultaneously. A video demonstrating the exploit is available from
the link\footnote{Kasa exploit: \kasavideo{}}.

%% file: exploit_lifx.tex
\subsubsection{Exploiting LIFX}\label{lifx-exploit}

LIFX~\cite{apk-lifx-playstore} controls
smart lights manufactured by LIFX, a company specialized
in smart lights~\cite{lifx}. The proof-of-concept exploit we developed is
a script that takes control of a light bulb. To develop the script,
we first started with finding answers to the four questions
in Section~\ref{sec:questions}.

\qone{}~We found that no encryption or
authentication is used in the LIFX app.

\qtwo{}~The LIFX app uses UDP to communicate with the smart lights. As in the Kasa case, we used the output of
the protocol discovery component to find that.

\qthree{}~The app uses broadcast messages. Listing~\ref{lst:lifx-broadcast} shows the class
responsible for sending these messages. Lines 5 and 6 refer to a fixed
broadcast address. We also found that the broadcast IP address was
identical to that present in the Kasa app. During our analysis of the
apps (discussed in Section~\ref{sec:discussion}), we noticed that this
broadcast address is commonly used in an IoT setup, as it represents a
special broadcast address that is used when a device needs to send a
broadcast packet to the network without caring about a recipient's
address.

\begin{lstlisting}[language=Java,
                 label={lst:lifx-broadcast},
                 caption={Function (simplified) used by the LIFX app
                 showing usage of a broadcast address.}
                ]
public final class UdpTransport ... { ...
  static {
    Object bn = InetAddress.getByName("255.255.255.255");
    Intrinsics.a(bn,...);
    ...
  } 
  ...
}
\end{lstlisting}

\qfour{}~During our analysis, we did not find any usages of protocols
with known vulnerabilities in the LIFX app.

\begin{figure}[h!]
  \centering
  \includegraphics[width=0.35\textwidth]{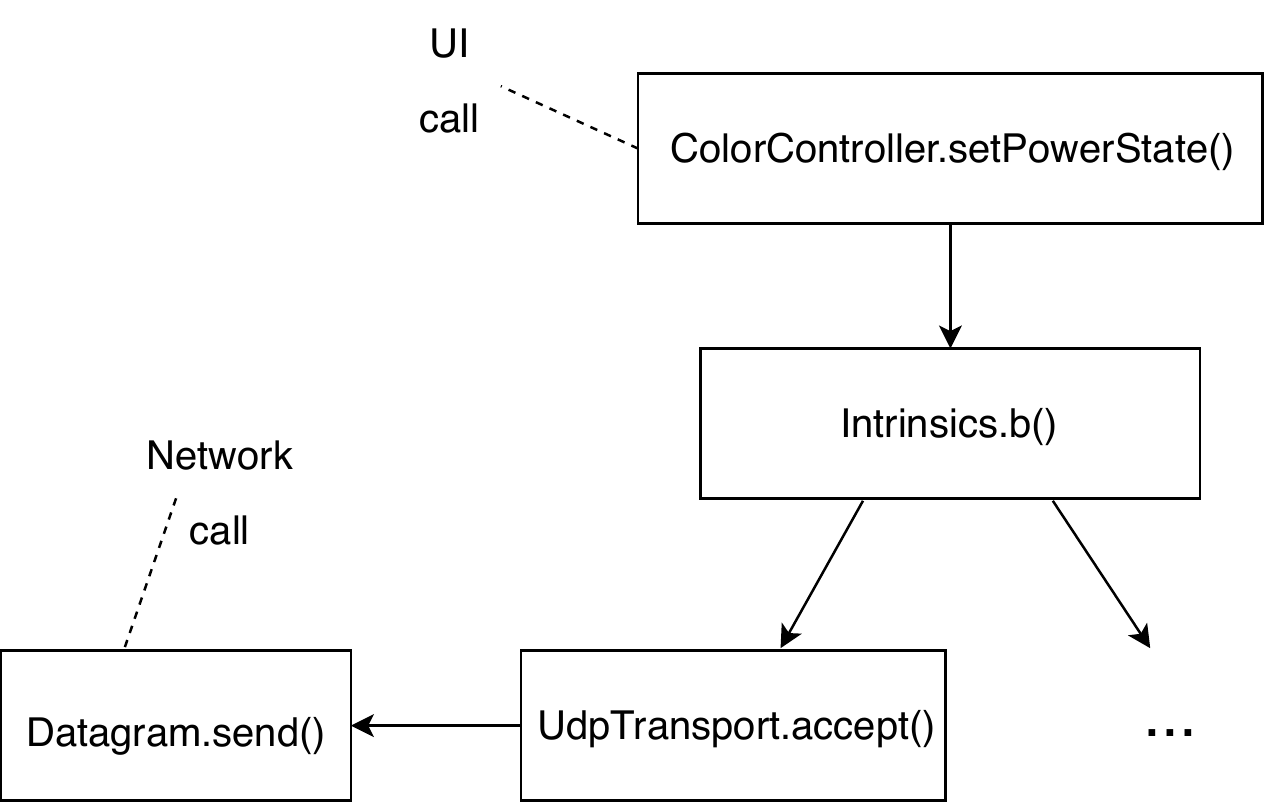}
  \caption{\label{fig:call-lifx}Path (simplified) from UI function to
  a network call. The path includes no encryption function.}
\end{figure}



The LIFX exploit consists of a script that sends UDP broadcast
messages to control the device. Similarly to the Kasa exploit, we used static
analysis to find a vulnerable attack vector for the app. For that, we inspected
the code and discovered the \CodeIn{UdpTransport} class,
responsible for creating and sending the UDP messages~(related the Q2). While inspecting this class, we found
that broadcast UDP messages were sent through the
\CodeIn{UdpTransport.accept} method (related to Q3). Then, we
analyzed the call chain backwards leading to the UI function responsible for
controlling the state of the smart light, \ie{}, turning it ``on'' and
``off''--\CodeIn{ColorController.setPowerState}, the starting point of
this path. Figure~\ref{fig:call-lifx} shows this path.

By monitoring the network while using the LIFX app to turn the
lights ``on'' and ``off'', we confirmed that broadcast messages were
exchanged during communication between the app and the device. Then, we proceeded to inspect the
code and find the structure of the
messages. We started from the network call from Figure~\ref{fig:call-lifx},
whose code is depicted in
Listing~\ref{lst:lifx-message-creation}. It is worth noting that
contrary to the Kasa app, which used a JSON structure for the message,
we could not infer the layout of the message by monitoring the network
only, as the message was in a byte array format.

\begin{lstlisting}[language=Java,
                 label={lst:lifx-message-creation},
                 caption={Code used by the LIFX app to construct an UDP message.}
                ]
class UdpTransport ... { ...
  void accept(@\textbf{TargetedMessage}@ targetedMsg) {@\label{line-targetedmessage}@
    Message msg = targetedMsg.getMessage();
    byte[] ba = msg.toData();
    datagramSocket.send(ba);
  } 
... 
}
\end{lstlisting}

\sloppy
We then inspected the parameter(s) of this method, \CodeIn{TargetedMessage}, highlighted in
Line~\ref{line-targetedmessage}. We found that the array of bytes
encapsulated on \CodeIn{TargetedMessage} objects is partly constructed within
classes encoding different operations. For example, the
class \CodeIn{SetPower} creates a message to modify the light power
whereas the class \CodeIn{SetColor} creates a message to modify the
light color. For illustration, Listing~\ref{lst:lifx-set-power}
shows a fragment of the class \CodeIn{SetPower}. The layout of messages can be inferred
from the method \CodeIn{put}. (The construction of the message header is omitted for brevity.)
The sequence of invocations to method \CodeIn{ByteBuffer.put*}
reflects the order of fields in the corresponding message. From that, we were able to
replicate the message layout in our script. We used the same rationale
to find the layout for other kinds of messages.

\begin{lstlisting}[language=Java,
                 label={lst:lifx-set-power},
                 caption={Structure of messages used to turn LIFX
                   devices on and off.}
]
class SetPower ... { ...
  final int SIZE = 6; long duration;  int level;
  void put(ByteBuffer byteBuffer) {
    byteBuffer.putShort(this.level);
    byteBuffer.putInt(this.duration);
  } 
... 
}
\end{lstlisting}


To summarize, we created a proof-of-concept exploit script to arbitrarily control LIFX smart
lights. Similar to the Kasa exploit described in Section~\ref{kasa-exploit}, we
used static analysis to find a vulnerable path and dynamic analysis to
understand the communication protocol between the LIFX app and
the smart light. To our surprise, we found that the app
did not implement pairing or encryption. Consequently, our script
sends unecrypted commands to change the state of the smart light, \eg, turning
it ``on'' and ``off'' and to change the color and saturation of the light.

%% file: exploit_wemo.tex
\subsubsection{Exploiting WeMo}\label{wemo-exploit}

WeMo is the official app to control Belkin devices from the product
line WeMo~\cite{wemo}, focusing on a variety of \iot\ devices in the
smart home segment. The proof-of-concept exploit we developed is a
script that arbitrarily controls a smart plug. As with earlier apps,
we started our analysis by answering the four questions from
Section~\ref{sec:questions}.

\qone{}~The WeMo app uses no encryption or authentication in
communication with the device.

\qtwo{}~The WeMo app uses the Universal Plug-and-Play (UPnP)
protocol to handle both device and service discovery on the local
network, with UDP acting as the underlying network protocol.

\qthree{}~Using UPnP implies that the SSDP protocol is used on the
app. As the SSDP protocol operates with a specific multicast
address--\CodeIn{239.255.255.250}-- broadcast messages are not used.

\qfour{}~We found that the app uses the UPnP
protocol. Although simple by design, this protocol is known to be
vulnerable. In 2013, a security company discovered that over 80 million devices were
susceptible to a UPnP vulnerability caused by excessive privilege in
the network interface of those devices~\cite{rapid7-upnp}. They found
that this vulnerability was actively being used in distributed
denial-of-service (DDoS) attacks.\vspace{1ex}

The WeMo exploit consists of a script that searches for a Belkin smart plug on the
network and sends commands for controlling it. To find an
attack path, we followed a different approach from previous
exploits.  As no encryption was present, we first monitored
the network to identify the structure of the unencrypted
messages. Then, by analyzing the code responsible for producing the message, we gathered the necessary information to build the script.

After identifying that the app uses the UPnP protocol, we
first tried to locate classes containing UPnP service names exposed by
the devices (related to Q2). These names
identify the universal device name (udn)~\cite{upnp-architecture}.
By design, they must start with the prefix ``\CodeIn{urn:}''. We found that the
\CodeIn{WeMoDevice} class contained the names of all services related
to the Belkin IoT devices, as seen in Listing~\ref{lst:wemo-devices}. This
is important as these names are part of the request sent to the
device. 

\begin{lstlisting}[language=Java,
                 label={lst:wemo-devices},
                 caption={Service names of WeMo devices complete list
                 was hidden for brevity.}
]
public class WeMoDevice { ...
  AIR_PURIFIER = "urn:Belkin:device:AirPurifier:1";
  LIGHT_SWITCH = "urn:Belkin:device:lightswitch:1";
  SWITCH = "urn:Belkin:device:controllee:1";
  public void setState(String state) { @\label{line-setstate}@
    this.mState = state;
  }
...
}
\end{lstlisting}\vspace{-1.5ex}

We then monitored the network while interacting with the WeMo
app\Comment{ to control the smart plug}. By monitoring the packets exchanged and
analyzing the response, we
observed that the app specifies a function---\CodeIn{SetBinaryState}---to change the state of the device and another
function---\CodeIn{GetBinaryState}---to obtain the state of the
device. Listing~\ref{lst:wemo-response} shows the response of the WeMo
plug while using the app. The response is a SOAP
envelope---the type of message used in the SSDP protocol.

\begin{lstlisting}[language=XML,
   label={lst:wemo-response}, caption={WeMo simplified response to query the
     current device state.}]
<s:Envelope ...>
    <s:Body>
        <u:GetBinaryStateResponse ...>
            <BinaryState>1</BinaryState>
        </u:GetBinaryStateResponse>
    </s:Body>
</s:Envelope>
\end{lstlisting}

While inspecting the code looking for the method
\CodeIn{SetBinaryState}, present in the WeMo response, we found the class \CodeIn{DeviceListManager}
that receives as parameters the state and udn of the device and calls
the
function \CodeIn{makeStateChangeRequest}. Listing~\ref{lst:wemo-devicelistmanager}
shows a fragment of class \CodeIn{DeviceListManager}{}. We then analyzed backwards the
call chain from method \CodeIn{makeStateChangeRequest} leading to the UI function responsible to turn the
smart plug ``on''.. The final vulnerable path was
hidden for brevity.

\begin{lstlisting}[language=Java,
                 label={lst:wemo-devicelistmanager},
                 caption={Class (simplified) responsible for sending
                   the command to change the state of a WeMo device.}
]
public class DeviceListManager ... { ...
  void setBinaryState(String state, String udn) { ...
    DeviceInformation device = getDeviceUDN(udn,...);
    makeStateChangeRequest(udn, device.getMAC(),...);
  }
  ...
}
\end{lstlisting}

To summarize, we successfully created a proof-of-concept exploit that
discovers WeMo devices compatible with the SSDP protocol located on
the local network and then can execute arbitrary commands on them. We
found that the app uses the UPnP protocol with well documented
vulnerabilities (see Table~\ref{table:cve}), but we did not end up
needing to use those vulnerabilities in the exploit since the devices
also have other vulnerabilities such as not using encryption.

%% file: exploit_econtrol.tex
\subsubsection{Exploiting e-Control}\label{econtrol-exploit}

e-Control is the app~\cite{apk-eControl-playstore} responsible for
controlling all devices from Broadlink---a company
specialized in smart home devices and universal remote
controllers. Our exploit was implemented as a script that
can arbitrarily control a Broadlink InfraRed (IR) remote controller. Similar to
previous exploits, we start by answering the questions of
Section~\ref{sec:questions}.

\qone{}~We did not find use of encryption or authentication in the app.

\qtwo{}~With our protocol discovery component, we found that the
e-Control app uses the QUIC UDP protocol to communicate locally with
Broadlink devices.

\qthree{}~We found usage of a global broadcast address (similar to
the Kasa app), implying the use of broadcast messages. The obfuscated
method at Line~\ref{line-broadcast-econtrol} of Listing~\ref{lst:econtrol-udp} shows this usage.

\qfour{}~We did not find usage of protocols with well-known
vulnerabilities in the app.

The e-Control exploit we developed consists of a script that can send
arbitrary commands to a Broadlink IR remote
controller. To create the script, we used the same methodology
as the LIFX exploit, using static analysis to
find a vulnerable path in the app. First, we inspected the code
looking for usages of the UDP protocol, where we found the
obfuscated method \CodeIn{a} of the class \CodeIn{PutInDataUnit}, responsible for sending the UDP packet
to the broadcast message (related to Q2 and
Q3). We then analyzed
usages of this method. We discovered that after scanning for devices
on the local network, the function \CodeIn{b} of the same class was
responsible for parsing the device's response (structured as a JSON
message). Listing~\ref{lst:econtrol-udp} shows the class with both
methods. We then analyzed backwards the call chain involving this
class and found the UI call, representing the start of the vulnerable
path. We choose not to show the graph illustrating the vulnerable path
for brevity reasons.

\begin{lstlisting}[language=Java,
                 label={lst:econtrol-udp},
                 caption={Class (simplified) responsible for sending an UDP message
                   through the network.}
                ]
public class PutInDataUnit {
  private Void a() { @\label{line-broadcast-econtrol}@...
    e = new DatagramSocket();
    e.send(new DatagramPacket(...,"255.255.255.255"));
    ...
  }

  private Void b(String... strArr) { ...
    JSONObject jSONObject = jSONArray.getJSONObject(i);
    ManageDevice device = new ManageDevice();
    device.setDeviceName(jSONObject.getString(a.d));
    ...
  }
}
\end{lstlisting}

To discover the structure of the messages, we  monitored the network
while interacting with the e-Control app. By inspecting the JSON
messages exchanged, we confirmed the structure of the message found
during our static analysis. We then replicated this structure to
control the device. It is worth noting that although we found the
functions related to the pairing process while inspecting the code, we
did not need to replicate them on our script. A Broadlink device can
be controlled independently of a previous successful pairing and without
alerting the user. This is a severe flaw.

%% file: discussion.tex
\begin{figure*}[ht]
  \begin{subfigure}{1\columnwidth}
    \centering
    \begin{minipage}[b]{1\textwidth}
      \input{pie-charts/hardcodedkeys}
      \caption{Apps regarding encryption.}\label{fig:piechart-hardcoded}
    \end{minipage}
  \end{subfigure}\hspace{2ex}
  \begin{subfigure}{1\columnwidth}
    \centering
    \begin{minipage}[b]{1\textwidth}
      \input{pie-charts/local}
      \caption{Apps regarding local communication.}\label{fig:piechart-local}
    \end{minipage}
  \end{subfigure}\\[1em]
  \begin{subfigure}{1\columnwidth}
    \centering
    \begin{minipage}[b]{1\textwidth}
      \input{pie-charts/broadcast}
      \caption{Apps regarding broadcast messages.}\label{fig:piechart-broadcast}
    \end{minipage}
  \end{subfigure}\hspace{2ex}
  \begin{subfigure}{1\columnwidth}
    \centering
    \begin{minipage}[b]{1\textwidth}
      \input{pie-charts/safe-protocol}
      \caption{Apps regarding secure protocols.}\label{fig:piechart-safe-protocol}
    \end{minipage}
  \end{subfigure}
  \caption{\label{fig:pie-charts}Distributions of different features for the set of analyzed apps.}
\end{figure*}
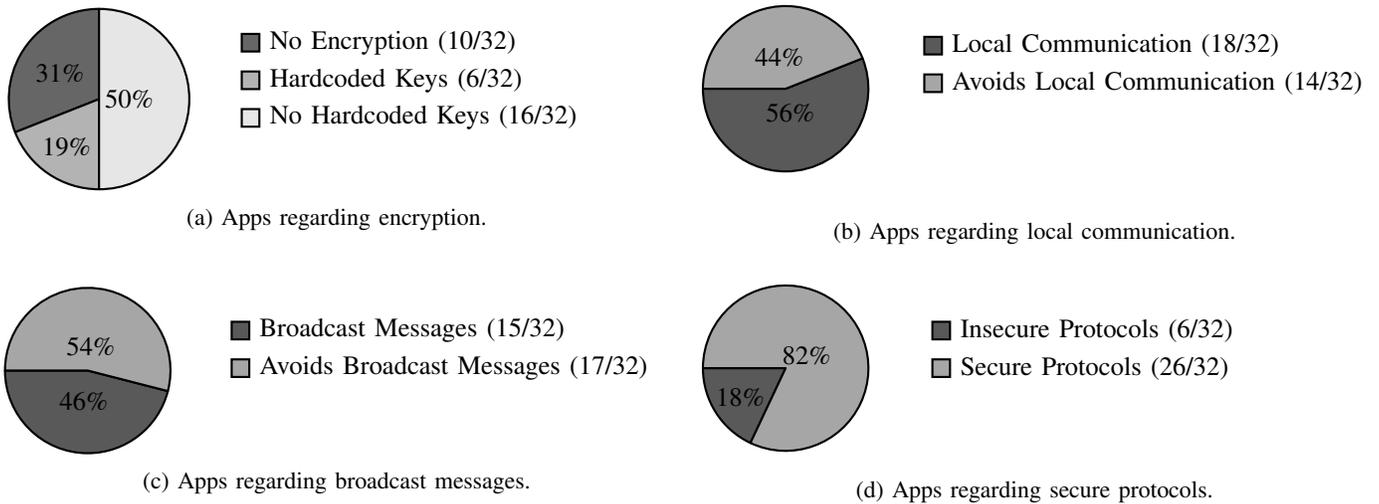

\section{Discussion}\label{sec:discussion}\label{subsec:step1}\label{sec:find-exploitables}


This section describes findings from a larger set of apps compared to the one used in Section~\ref{sec:finding-confirming} that enabled us to construct exploits. In this case,
we did not \emph{not} purchase devices, so these findings are indicative of
the potential extent of vulnerabilities. IoT devices included cameras, locks,
and alarms, suggesting a significant safety issue if the vulnerabilities we
found were exploitable. We consider companion app analysis
to be a valuable tool for discovering potential vulnerabilities prior to
making purchases of large or expensive smart appliances or where human safety
could be an issue. Our analysis is
conservative with respect to potential vulnerabilities.
We only present a negative result if we
can confirm it in code.
For example, by analyzing the eWeLink app, which uses native code, our
analysis was not helpful in analyzing the app for hardcoded keys and, to be
conservative, we counted the app as not having vulnerabilities to hardcoded
keys.





For this experiment, we start with the top-100 smart hubless devices on
the Amazon website by popularity and then restrict the resulting set to
devices that use WiFi and Bluetooth for communication. We find 96
such devices, including the 5 devices that we previously purchased and analyzed
in Section~\ref{sec:threat-model}. These \totDevices{} devices
only correspond to \totApps{} companion apps, saving us significant analysis effort compared to analysis of devices themselves.


Figure~\ref{fig:pie-charts} shows the distribution of answers to the questions for
the analyzed apps as pie charts. To answer these questions we used the same
methodology described in Section~\ref{sec:threat-model}.
Considering these questions, we find 31\% of the apps to not use any encryption at all and 19\% to use hardcoded keys, strongly suggesting that at least 50\% of the apps are potentially exploitable with further protocol analysis.
These correspond to 37 out of 96 devices, i.e., 38\% of the devices.

Out of the 32 apps analyzed, we found 4 apps to use encryption without
hardcoded keys, not use local communication, not use broadcasts, and
not use known insecure protocols.
All their communication was via the cloud service, likely over SSL\@.
The four apps include the popular Nest app. With respect to attacks considered
in this paper, this is a relatively secure way to communicate.
But it does have a privacy tradeoff in that the cloud service has
access to the commands sent to the device. Consequently, a potential
long-range security risk exists if
the cloud service is ever compromised, a non-negligible risk~\cite{dtap18}.

We observe that \localComm{}
apps communicate locally with their corresponding device(s) and that
\broadcast{} apps use broadcasting. Note that a negative answer to a question does not imply an exploitable vulnerability.
Further inspection is necessary to understand the
issue. For example, although the app ``August Home'' uses local
communication, that happens only through Bluetooth, which may restrict an
attacker to be in close proximity to the device, limiting the attack
surface. The app
``Ring---Always Home'' also does not appear problematic. After closer
inspection, we find that the use of SIP, a protocol known to be
insecure (see Table~\ref{table:cve}) is due to one specific feature of
the app, Voice over IP (VoIP). However, this feature can be disabled
and, if enabled, it is isolated from other parts of the app (with no
apparent flows in between).

%% file: pie-charts/hardcodedkeys.tex
\begin{tikzpicture}  
  \pie [text=legend,rotate = 90,radius=1.2,
    color={black!60,
       black!30, black!10}]
       {31/No Encryption (10{/}32),
         19/Hardcoded Keys (6{/}32),
         50/No Hardcoded Keys (16{/}32)}
\end{tikzpicture}

%% file: pie-charts/local.tex
\begin{tikzpicture}
  \pie [text=legend,rotate = 180,radius=1.1,
  color={black!65,black!35}]
    {56/Local Communication (18{/}32),
     44/Avoids Local Communication (14{/}32)}
\end{tikzpicture}

%% file: pie-charts/broadcast.tex
\begin{tikzpicture}  
  \pie [text=legend,rotate = 180,radius=1.1,
    color={black!65,black!35}]
    {46/Broadcast Messages (15{/}32),
     54/Avoids Broadcast Messages (17{/}32)}
\end{tikzpicture}

%% file: pie-charts/safe-protocol.tex
\begin{tikzpicture}
  \pie [text=legend,rotate = 180,radius=1.1,
  color={black!65, black!35}]
    {18/Insecure Protocols (6{/}32),
     82/Secure Protocols (26{/}32)}
\end{tikzpicture}

%% file: defense.tex
\section{Defense Strategies}\label{sec:lessons}
Our analysis suggests that developers of software for IoT devices find
it non-trivial to do proper key management. 50\% of the apps either
used hardcoded keys or did not do encryption. Some apps though did not
use hardcoded keys and it is instructive to see how they secured
app-device communication.

We have not found use of a hardcoded key or of local communication in Nest thermostat's companion app.
The Nest thermostat provides a UI interface that provides a display and also entry of data (by rotating a ring around the thermostat). A user can use that interface that to select a WiFi network and enter the WiFi password.
That allows the device to talk to the Nest cloud service securely over SSL\@. The companion app does not talk directly to the device; instead, the user creates a free account on the Next cloud service and then signs into that using the companion app over SSL\@. Then, the user adds the thermostat to his account by entering a random code
that is displayed on the thermostat's interface. The user that way is assured that the correct thermostat is being added to
the account. Furthermore, the thermostat and the cloud service can also mutually authenticate each other
and establish a shared secure link.  No shared keys between the
companion app and the thermostat are required since, from then on, the
communication between the companion app and the thermostat happens over SSL links to
the cloud service.

 The EZVIZ uses a different strategy. Unlike Nest, it supports
 local communication between the companion app and the device over the
 local network. The shared encryption key is enclosed in the box in
 the form of a QR code and must be scanned by the companion app.  This
 strategy is better than hardcoded keys provided the key in the QR
 code is of sufficient length, random, and strong crypto library is
 used.

 Certain strategies are not recommended. We found several apps rely on native code.
 In our current analysis, we did not analyze native code and thus could have missed some uses of hardcoded keys.
 But, it is certainly possible to analyze binary code using tools such as Ida Pro~\cite{ida-pro} and extract potential
 constants, including keys. Similarly, code obfuscation, is unlikely to be an adequate defense.


%% file: related.tex
\section{Related Work}\label{sec:related}

In this section, we review some of the previous studies on security in
a smart home context.


Denning \etal{}~\cite{denning-kohno-levy-2013} presented potential
security attacks against smart home devices, pointing that common
attacks to traditional computing platforms, like denial-of-service and
eavesdropping on network could also be used in a smart home
context. Komninos \etal{}~\cite{komninos-philippou-pitsillides-2014} presented
a survey about smart home security, pointing different usage scenarios
and categorizing threats into network domain (such as eavesdroping,
traffic analysis and replay attack) and smart home introduced
concept (such as device impersonation,
update and illegal software).


Focusing on IoT platforms, Fernandes \etal{}~\cite{fernandes-jung-prakash-2016}
analyzed over 499 apps on SmartThings and found out that 55\% of those are
over-privileged largely due to design flaws in the privilege model of the platform. The authors
also demonstrated how to take advantage of this with four
proof-of-concept attacks, both remotely and locally. Jia \etal{}
proposed a context-based permission system for appified IoT platforms with
fine-grained context identification and runtime prompts~\cite{contexiot}.
Other works~\cite{soteria, pasurvey} focus on program analysis techniques used
on IoT platforms. Most of these works analyze apps written in platform-dependent
and restricted languages whereas we analyze companion apps in the larger Android platform.




Android apps have been analyzed for a variety of security-related issues, such as
cryptographic misuse~\cite{crypto-api-misuse, amandroid, javacrypto} and memory corruption~\cite{iotfuzzer}.
For example, Egele~\etal{}~\cite{crypto-api-misuse} analyzed the violation of six rules including the
use of ECB mode and constant keys/IVs/seeds.
Wei \etal{}~\cite{amandroid} designed a static analysis tool for security vetting of
Android apps and used it to detect the use of the weak ECB mode for encryption;
the analysis is intra-procedural and thus limited in scope.


  In 2015, the Veracode team published a white paper on security analysis of
  six IoT devices to examine vulnerabilities involving non-use of cryptography, lack of enforcement of  strong passwords, and incorrect TLS certificate validation~\cite{veracode-whitepaper-2015}. The white paper mentions that the team used   network monitoring and reverse engineering techniques, but did not discuss
  details of what was reverse-engineered (e.g., device firmware, apps, or cloud services) or how.  Our work differs in that it focuses on a different 
  set of vulnerabilities (e.g., many of the apps we analyzed in 2018 use
  cryptography, but with hard-coded keys) and we present details of analysis
  of companion apps to show how such vulnerabilities can be discovered.


%% file: conclusion.tex
\section{Conclusions}\label{sec:conclusions}

Securing communication between \iot\ devices and the mobile apps
responsible for controlling them is crucial for security and even
safety, depending on the types of IoT devices on a
network. Unfortunately, analysis of device firmware is usually
non-trivial.  In this study, we showed that analyzing the smartphone
companion apps that are released for the device can provide important
clues for potential vulnerabilities in the devices. By analyzing
the companion app code, we assessed whether the communication between five
best-selling \iot\ devices and their companion apps occurs over a
secure channel. We found that was not the case. We were successful
in creating exploits for all five devices and able to
control them, leveraging information that we gathered
while analyzing the companion apps, both statically, through program
analysis, and dynamically, through monitoring the network. We also
extended our study to 28 additional apps. We found that 31\% of the apps
do not use any crypto to protect the device-app communication and that 19\%
use hardcoded keys. A significant fraction of the apps (40--60\%) also
use local communication or local broadcast communication, thus providing
an attack path to exploit lack of crypto or use of hardcoded encryption keys.
